\documentclass[epj]{svjour}
\usepackage{latexsym}
\usepackage{amsmath,amssymb,amsfonts}
\usepackage{graphics}

\usepackage{epsf}

\begin{document}
\title{A quantum trajectory description of decoherence}
\author{A. S. Sanz\inst{1} \and F. Borondo\inst{2}}
%
%
\institute{Instituto de Matem\'aticas y F\'{\i}sica Fundamental,
Consejo Superior de Investigaciones Cient\'{\i}ficas, \\
Serrano 123, 28006 Madrid, Spain \and
Departamento de Qu\'\i mica, C--IX, Universidad Aut\'onoma de Madrid,
Cantoblanco -- 28049 Madrid, Spain}
%
\date{}
%
\abstract{
A complete theoretical treatment in many problems relevant to
physics, chemistry, and biology requires considering the action
of the environment over the system of interest.
Usually the environment involves a relatively large number of degrees
of freedom, this making the problem numerically intractable from a
purely quantum--mechanical point of view.
To overcome this drawback, a new class of quantum trajectories is
proposed.
These trajectories, based on the same grounds as Bohmian ones, are
solely associated to the system reduced density matrix, since the
evolution of the environment degrees of freedom is not considered
explicitly.
Within this approach, environment effects come into play through a
time--dependent damping factor that appears in the system equations
of motion.
Apart from their evident computational advantage, this type of
trajectories also results very insightful to understand the system
decoherence.
In particular, here we show the usefulness of these trajectories
analyzing decoherence effects in interference phenomena, taking as
a working model the well--known double--slit experiment.
} 

\maketitle


\section{Introduction}
\label{sec1}

Among the different alternative mechanisms proposed to explain how the
behavior of a quantum system becomes classical--like \cite{Omnes1},
decoherence is the most widely accepted \cite{Zurek1,Giulini1}.
Decoherence is the irreversible emergence of classical properties
when an isolated system, namely the system of interest, interacts
with an environment \cite{Kiefer1}.
The environment can be constituted by many randomly distributed
particles interacting with the system by means of scattering processes.
When these events occur in a large number, the off--diagonal elements
of the system reduced density matrix undergo an exponential damping
\cite{Joos1}, this making the system to quickly lose its coherence.
Coherence loss is an important issue, for example, in quantum
computation \cite{Nielsen}, where long chains of atoms must be kept in
a coherent superposition for certain time in order to perform the
corresponding operations.
Therefore, decoherence effects must be considered seriously; they
increase rapidly with the length of the chain \cite{Unruh}, thus
decreasing the efficiency of the latter in performing such operations.

At a first glance, Bohmian mechanics (BM) \cite{Bohm,Holland} seems to
be a suitable tool to study and shed light on decoherence problems.
Unlike the standard version of quantum mechanics (SQM), the Bohmian
formalism is based on the concept of well--defined trajectories;
particles are always regarded as particles, as in classical mechanics.
The particle motion, governed by the wave function, leads to the same
results provided by SQM when a sampling over particle initial
conditions is considered (see, for example, reference~\cite{Sanz0}).
This capability to conjugate motion and statistical predictions within
a purely quantum framework has been widely used to describe, for
example, interference experiments with slits
\cite{Philippidis,Wiseman2,Ghose1,Sanz1,Sanz2,Marchildon1}, where
decoherence can play an important role.

Nevertheless, despite the apparent suitability of BM to the study of
decoherence, in practice its application results numerically
prohibitive in problems where the number of degrees of freedom
involved becomes relatively large.
To overcome this computational drawback one can proceed as in SQM when
dealing with Markovian environments \cite{Breuer}.
In these cases, decoherence effects can be studied by using a master
equation formulation, where the particular time--evolution of the
different environment degrees of freedom is not taken into account
explicitly.
The Markovian master equation, derived from the von Neumann equation
for the whole system (the system of interest plus the environment),
is generally expressed as a sum of two contributions, which are
responsible for: (1) the time--evolution of the isolated system, and
(2) the quenching leading to the system coherence loss.
All the information regarding the physical properties of the
environment as well as its interaction with the system is contained
within this second term.

Starting from BM, one can proceed as in SQM and consider the
``average'' action of the environment degrees of freedom in order
to obtain a Markovian--like trajectory equation of motion.
We can thus define a {\it new} class of quantum trajectories, the
{\it reduced quantum trajectories}, directly related to the system
reduced density matrix, but also influenced by the presence of the
environment.
Within this formulation, the environment degrees of freedom come into
play through a time--dependent damping factor that appears in the
expression for the particle velocity field.
Avoiding to integrate explicitly the equations of motion for the
environment degrees of freedom allows to obtain an important insight
on the system dynamics at a low computational cost.

To illustrate the applicability and interest of the reduced quantum
trajectories, we will use them to analyze the effects of decoherence
in interference phenomena.
These phenomena constitute an ideal framework to study decoherence
because of their simplicity as well as their fundamental implications
in quantum mechanics.
In particular, we will focus our discussion on the double--slit
experiment, which can be regarded as the paradigm of quantum
interference.
In this experiment, in the absence of ``which--way'' information, the
measured intensity displays the well--known interference pattern with
maximum fringe visibility.
On the other hand, the knowledge of the particle pathway destroys
such a pattern and the intensity thus acquires classical features,
i.e., it is simply given by the sum of the intensities corresponding
to each pathway\footnote{Except otherwise stated, throughout this work
``classical'' refers to the lack of quantum interference, though in
general this does not mean necessarily lack of other quantum effects
(e.g., single--slit diffraction \cite{Sanz1}).}.
These two possible outcomes\footnote{Though in SQM textbooks these two
situations are mutually exclusive, it has been shown both theoretically
\cite{Wootters} and experimentally \cite{SDuerr} that it is still
possible to determine certain amount of ``which--way'' information
without a full erasure of the interference pattern.}, related to which
aspect of the particle we are interested in (wave or corpuscle,
respectively), are equivalent to consider two different experimental
contexts: both slits simultaneously open or each one independently open.
The choice of a quantum context determines the intensity pattern, in
sharp contrast to what happens in classical mechanics in the analogous
situation, where both contexts give the same result.
However, if the action of an external environment (air molecules,
thermal photons, etc.) over the system is taken into account between the
slits and the detector, a partial (or even total) suppression of the
quantum interference will be observed in the intensity pattern.
This means that a certain amount of ``which--way'' information is being
gradually revealed, and the process can be thought as a smooth
transition from the context where both slits are simultaneously open
to the other one where they are independently open.

The organization of this paper is as follows.
In Section~\ref{sec2} we introduce the formal grounds of the reduced
quantum trajectory formalism as well as its theoretical application
to the double--slit problem.
In Section~\ref{sec3} we present an application of this formalism to
the double--slit experiment with cold neutrons performed by Zeilinger
{\it et al.} \cite{Zeilinger1}.
Finally, in Section~\ref{sec4} the main conclusions arisen from this
work are summarized.


\section{Decoherence and quantum trajectories}
\label{sec2}


\subsection{The reduced quantum trajectory approach}

In order to extract useful information about the system of interest,
one usually computes its associated reduced density matrix by tracing
the total density matrix\footnote{Throughout this work, we indicate
time--dependence by a subscript ``$t$'' (e.g., $\hat{\rho}_t \equiv
\hat{\rho} (t)$), while initial values do not carry any label (e.g.,
$\hat{\rho} \equiv \hat{\rho} (0)$).}, $\hat{\rho}_t$, over the
environment degrees of freedom.
In the configuration representation and for an environment constituted
by $N$ particles, the system reduced density matrix is obtained after
integrating $\hat{\rho}_t \equiv |\Psi\rangle_t\ \! _t\langle\Psi|$
over the 3$N$ environment degrees of freedom, $\{ {\bf r}_i \}_{i = 1}^N$,
\begin{eqnarray}
 \tilde{\rho}_t ({\bf r}, {\bf r}') & = &
  \int \langle {\bf r}, {\bf r}_1, {\bf r}_2, \ldots {\bf r}_n |
  \Psi \rangle_t
 \nonumber \\
 & & \times\ _t\langle \Psi | {\bf r}', {\bf r}_1, {\bf r}_2,
  \ldots {\bf r}_n \rangle \ \! {\rm d}{\bf r}_1 {\rm d}{\bf r}_2
  \cdots {\rm d}{\bf r}_n .
 \label{eq7}
\end{eqnarray}
The system (reduced) quantum density current can be derived from this
expression \cite{Appleby,Viale}, being
\begin{equation}
 \tilde{\bf J}_t \equiv \frac{\hbar}{m}
  \ {\rm Im} [ \nabla_{\bf r} \tilde{\rho}_t ({\bf r},{\bf r}')]
  \Big\arrowvert_{{\bf r}' = {\bf r}} ,
 \label{eq12}
\end{equation}
which satisfies the continuity equation
\begin{equation}
 \dot{\tilde{\rho}}_t + \nabla \tilde{\bf J}_t = 0 .
 \label{eq13}
\end{equation}
In equation (\ref{eq13}), $\tilde{\rho}_t$ is the diagonal element
(i.e., $\tilde{\rho}_t \equiv \tilde{\rho}_t ({\bf r},{\bf r})$) of the
reduced density matrix and gives the measured intensity \cite{Sanz3}.

Taking into account equations (\ref{eq12}) and (\ref{eq13}), now we
define the velocity field, $\dot{\bf r}$, associated to the (reduced)
system dynamics as
\begin{equation}
 \tilde{\bf J}_t = \tilde{\rho}_t \dot{\bf r} ,
 \label{vfield}
\end{equation}
which is analogous to the Bohmian velocity field.
Now, from equation (\ref{vfield}), we define a new class of quantum
trajectories as the solutions to the equation of motion
\begin{equation}
 \dot{\bf r} \equiv \frac{\hbar}{m}
  \frac{{\rm Im} [ \nabla_{\bf r} \tilde{\rho}_t ({\bf r},{\bf r}')]}
       {{\rm Re} [ \tilde{\rho}_t ({\bf r},{\bf r}')]}
   \Bigg\arrowvert_{{\bf r}' = {\bf r}} .
 \label{eq14}
\end{equation}
These new trajectories are related to the system reduced density
matrix, therefore we call them the {\it reduced quantum trajectories}.
In Section~\ref{sec3} we will see that the dynamics described by
equation (\ref{eq14}) leads to the correct intensity (whose
time--evolution is described by equation (\ref{eq13})) when the
statistics of a large number of particles is considered.
Moreover, also observe that equation (\ref{eq14}) reduces to the
well--known expression for the velocity field in BM when there is no
interaction with the environment.
This can be shown as follows.
The decoupling from the environment makes the system reduced density
matrix to be just the system density matrix.
Making then use of the BM ansatz for the system wave function
($\langle {\bf r}|\Psi\rangle_t = R_t ({\bf r}) \ \!
e^{i S_t ({\bf r}) /\hbar}$), we can express the system density
matrix as
\begin{equation}
 \rho_t ({\bf r},{\bf r}') =
  R_t R_t' \ \! {\rm e}^{i (S_t - S_t')/\hbar} ,
 \label{eq15}
\end{equation}
with $R'_t = R_t ({\bf r}')$ and $S_t' = S_t ({\bf r}')$.
Finally, substituting (\ref{eq15}) into equation (\ref{eq14}), one
reaches
\begin{equation}
 \dot {\bf r} = \frac{\nabla S_t}{m} ,
 \label{eq16}
\end{equation}
which, effectively, is the well--known expression for the particle
equation of motion in BM.

As mentioned above, BM becomes numerically in\-trac\-ta\-ble when the
phenomena described involve a large number of degrees of freedom.
Hence a wide range of alternative, approximate formulations rooted in
this approach have been proposed in the literature \cite{chapter}.
The reason behind formulating these alternative formalisms is quite
simple.
BM is the only trajectory--based approach compatible with SQM where
no approximations are considered, this leading to a straightforward
interpretation of quantum phenomena in terms of a self--consistent
quantum theory of motion (all the elements contained in the theory are
ruled by quantum laws).
Therefore any approximation to BM will also remain relatively close
to SQM, reducing at the same time the computational efforts implicated
by many degree--of--freedom systems.
As we have seen, these features also meet in our approach, whose
mathematical structure remains very close to that of BM.

Nevertheless, the aforementioned formulations are not so widespread as
those other based on the so--called semiclassical approximation of the
wave function, in particular, the semiclassical initial value
representation (SC--IVR) \cite{miller}, which is one of the most
successful approaches to date.
Unlike the previous prescriptions, the self--con\-sis\-ten\-cy
mentioned above breaks in SC--IVR schemes: on the one hand we have the
calculation of purely classical trajectories; on the other hand, there
is a (semiclassical) wave function which is calculated from those
classical trajectories.

The intertwining between classical and quantum mechanics in a
Feynman--like fashion \cite{Sanz2,feynman} constitutes the main
difference with respect to Bohmian--like schemes, such as the one
described in this work.
This difference can be noticed, for instance, when looking at the
dynamics displayed by the trajectories representative of each type of
formalism.
In semiclassical approaches trajectories will be just classical, not
showing any particular effect typical of the quantum problem treated
with them; only when these trajectories are introduced into the
semiclassical wave function a quantum description of the phenomenon
that we are dealing with can be obtained.
On the contrary, in Bohmian--like schemes, the trajectories display a
topology that is in accordance (in a more or less degree, depending on
the approximation considered) with the dynamics prescribed by quantum
laws.
In other words, even considering the same initial condition for both
types of trajectories the differences between the motions described by
each one will manifest immediately.
This comparison can be seen in reference \cite{Sanz2}, where
classical and Bohmian trajectories were obtained within the context
of the double--slit experiment with no coupling to an environment
(calculations applying the SC--IVR formalism to the double--slit
experiment can be seen in references (e) and (f) in \cite{miller},
for example, but no classical trajectories related were explicitly
shown).


\subsection{Reduced trajectory dynamics in the double--slit
experiment}

The implications and usefulness of equation~(\ref{eq14}) are better
appreciated when analyzing decoherence effects in the double--slit
experiment.
Quantum mechanically the evolution of a particle after passing through
a double--slit setup (and without being acted by an external environment)
can be described at any subsequent time by a wave function
\begin{equation}
 |\Psi^{(0)}\rangle_t = c_1 |\psi_1\rangle_t + c_2 |\psi_2\rangle_t ,
 \label{eq1}
\end{equation}
where $|\psi_j\rangle_t$ is the partial wave emerging from the slit
$j$ (with $j = 1, 2$), and $|c_1|^2 + |c_2|^2 = 1$ at any time.
In configuration space, the density matrix associated to the wave
function (\ref{eq1}) is
\begin{equation}
 \rho_t^{(0)} ({\bf r}, {\bf r}') =
  \Psi^{(0)}_t ({\bf r}) \left[ \Psi^{(0)}_t ({\bf r}') \right]^* ,
 \label{eq2}
\end{equation}
with $\Psi^{(0)}_t ({\bf r}) = \langle {\bf r} |\Psi^{(0)}\rangle_t$.
As said above, the diagonal of equation~(\ref{eq2}) gives the measured
intensity (or probability density),
\begin{eqnarray}
 \rho_t^{(0)} ({\bf r}) & = &
  |c_1|^2 |\psi_1|_t^2 + |c_2|^2 |\psi_2|_t^2
 \nonumber \\
 & & \qquad + 2 |c_1| |c_2| |\psi_1|_t |\psi_2|_t \cos \delta_t
 \label{eq3}
\end{eqnarray}
where $\delta_t$ is the time--dependent phase shift between the
partial waves.

Under the presence of an environment, the wave function (\ref{eq1})
does no longer describe the evolution of the isolated system.
To obtain an appropriate ansatz, first we consider elastic
system--environment scattering conditions, which lead to a gradual
suppression of the interference terms in equation~(\ref{eq3}) without
changing too much the states describing the system (i.e., each partial
wave).
Under these conditions, only the environment states, $|E_j\rangle_t$,
associated with each partial wave will change during the scattering
process.
In addition, we also assume that the system is initially represented by
a superposition of two Gaussian wave packets (see Section~\ref{sec3});
both partial waves will then keep their Gaussian shape during their
time--evolution, this simplifying the analysis of the problem.
Taking this into account, a general initial separable coherent state
\begin{equation}
 |\Psi \rangle = |\Psi^{(0)}\rangle \otimes |E_0\rangle
 \label{eq4}
\end{equation}
(with $|\Psi^{(0)}\rangle$ as in equation (\ref{eq1}) at $t=0$) will
become entangled,
\begin{equation}
 |\Psi \rangle_t =
    c_1 |\psi_1\rangle_t \otimes |E_1\rangle_t
  + c_2 |\psi_2\rangle_t \otimes |E_2\rangle_t ,
 \label{eq5}
\end{equation}
at any subsequent time.
Let us recall that the condition of initial coherence also means that
$|E_1\rangle = |E_2\rangle = |E_0\rangle$ at $t=0$.

Provided elastic system--environment scattering conditions as well as
weak interactions are assumed the decoherence process can be described
by means of (\ref{eq5}).
However, this assumption will no longer be valid as the
system--environment coupling becomes stronger and energy transfer
inelastic processes take place.
Such processes would imply a more complicated form for the wave
function, which could not be expressed in terms of two overlapping,
spreading Gaussians.
Instead, the energy transfer leads not only to a change of the shape
of the diffracted beams, but, more importantly, to a mixed state that
cannot be described in general as a simple wave function due to the
strong intertwining between the environment states and the system ones.
In such cases, one should to consider either the language of density
matrices commonly used in the theory of open quantum systems
\cite{Breuer,pechukas}, or the stochastic wave functions that appear
in the quantum state diffusion prescription \cite{percival}.

The effects induced by strong couplings can be better appreciated, for
instance, when using the SC--IVR formalism mentioned above, where the
environment is characterized by a certain spectral density of
frequencies (in general the environment is assumed to be a bath
harmonic oscillators with an Ohmic spectral density).
The aforementioned intertwining between environment and system states
leads to a very appealing loss of interferences in both bound systems
(e.g., vibrating diatoms in solvents; see reference (d) in
\cite{miller}) and problems in the continuum (e.g., inelastic
scattering of He atoms from a Cu surface; see reference (g) in
\cite{miller}).
Nonetheless, as the coupling strength increases more, this type of
descriptions may also lose their validity, since the bath of harmonic
oscillators will not longer describe properly the system--environment
coupling, which could be more complicated and lead to dissipation.

From (\ref{eq5}) the measured intensity is obtained from the system
reduced density matrix, which is given by the trace of the full density
matrix over the environment states,
\begin{equation}
 \hat{\tilde{{\rho}}}_t =
  \sum_{j=1}^2 \ _t\langle E_j | \hat{\rho}_t | E_j \rangle_t
 \label{eq6}
\end{equation}
(notice that this expression is equivalent to (\ref{eq7})).
Thus, substituting the wave function (\ref{eq5}) in equation
(\ref{eq6}) gives
\begin{eqnarray}
 \tilde{\rho}_t ({\bf r}, {\bf r}') =
 (1 + |\alpha_t|^2) \sum_{j=1}^2
  |c_j|^2 \psi_{j,t} ({\bf r}) \psi_{j,t}^* ({\bf r}')
  \nonumber \\
  + 2 \alpha_t c_1 c_2^* \psi_{1t} ({\bf r}) \psi_{2t}^* ({\bf r}')
  + c.c. ,
 \label{eq8}
\end{eqnarray}
where $\alpha_t \equiv \ \! _t\langle E_2 | E_1 \rangle_t$ and $c.c.$
means conjugate complex, and from this expression the measured
intensity results
\begin{eqnarray}
 \tilde{\rho}_t & = &
 (1 + |\alpha_t|^2) \left[ |c_1|^2 |\psi_1|_t^2 + |c_2|^2 |\psi_2|_t^2
  \right. \nonumber \\ & & \left. \qquad \qquad +
  2 \Lambda_t |c_1| |c_2| |\psi_1|_t |\psi_2|_t \cos \delta'_t \right] ,
 \label{eq9}
\end{eqnarray}
with
\begin{equation}
 \Lambda_t \equiv \frac{2 |\alpha_t|}{(1 + |\alpha_t|^2)}
 \label{eq10}
\end{equation}
being the {\it coherence degree}, which gives the value of the
{\it fringe visibility} in a good approximation \cite{Sanz3}.
For the sake of simplicity, we have assumed that the phase difference
between the environment states (included in $\delta'_t$) is constant.

The environment states are considered to be too complicated for keeping
mutual coherence as time increases \cite{Omnes2}; even if they are
initially coherent, they will become orthogonal along time.
Thus, one can assume $|\alpha_t| = {\rm e}^{-t/\tau_c}$, $\tau_c$
being the coherence time, a measure of how fast the system looses its
coherence.
Introducing this value into equation (\ref{eq10}), we obtain
\begin{equation}
 \Lambda_t = {\rm sech} (t/\tau_c) ,
 \label{eq11}
\end{equation}
which establishes a relationship between the coherence degree and the
coherence time.
Thus, although the value of the coherence time can be derived
analytically for interfering waves by means of simple Markovian models
\cite{Savage1}, equation (\ref{eq11}) allows us to determine it from
the empirical value of $\Lambda_t$ \cite{Sanz3} (i.e., measured from
the intensity pattern) and the time--of--flight, $t_f$, of the
diffracted particles.
Because of the empirical nature of $\tau_c$ (or, equivalently, the
value of $\Lambda_t$ after a full flight) in our model, temperature
does not appear explicitly despite its important role in decoherence
phenomena.
Within the context of the double--slit experiment, this issue has been
treated in some theoretical \cite{Savage1,qureshi} and computational
(references (e) and (f) in \cite{miller}) works in the literature.
In this way, following references \cite{Savage1,qureshi}, $\tau_c$ would
already encompass the effects of the temperature.
Note that this dependence on temperature is quite different from that
in semiclassical treatments; in the first case it is simply a parameter,
where in the latter it appears as a consequence of considering the bath
of oscillators at the equilibrium at a given temperature (i.e., the
oscillators follow a Boltzmann distribution).

As said above, equation (\ref{eq14}) has been defined in such a way
that there is no an explicit dependence on the dynamical evolution of
the environment degrees of freedom.
Nevertheless, the environment effects on the system will still be
present through $\alpha_t$, as seen when substituting (\ref{eq8}) into
equation (\ref{eq14}).
This can be seen in the resulting equation of motion
\begin{eqnarray}
 \dot{\bf r}_t & = & \frac{(1 + |\alpha_t|^2)\hbar}{2 i m
  \tilde{\rho}_t} \sum_{j=1}^2
   |c_j|^2 \left[ \ \! \psi_{j,t}^* \nabla \psi_{j,t}
    - \psi_{j,t} \nabla \psi_{j,t}^* \right]
 \nonumber \\
 & + & \frac{\hbar}{i m \tilde{\rho}_t} \ \!
  \alpha_t c_1 c_2^*
  \left[ \ \! \psi_{2,t}^* \nabla \psi_{1,t}
   - \psi_{1,t} \nabla \psi_{2,t}^* \ \! \right] + c.c.
 \label{eq17}
\end{eqnarray}

In the limit of total loss of coherence (i.e., $t \gg \tau_c$ or
$\alpha_t \rightarrow 0$), equation (\ref{eq17}) becomes
\begin{equation}
 \dot{\bf r}_t = \frac{|c_1|^2 \rho_t^{(1)} \dot{\bf r}_1 +
  |c_2|^2 \rho_t^{(2)} \dot{\bf r}_2}{\rho_t^{\rm cl}}
 \label{eq18}
\end{equation}
where $\dot{\bf r}_j$ and $\rho_t^{(j)}$ are, respectively, the
velocity field and the probability density associated to the partial
wave $|\psi_j\rangle_t$, and
\begin{equation}
 \rho_t^{\rm cl} \equiv |c_1|^2 \rho_t^{(1)} + |c_2|^2 \rho_t^{(2)} .
 \label{rhodec}
\end{equation}
Notice that both $\rho_t^{\rm cl}$ and the density current,
\begin{equation}
 {\bf J}_t^{\rm cl} \equiv \rho_t^{\rm cl} \ \dot{\bf r}_t =
  |c_1|^2 \rho_t^{(1)} \ \dot{\bf r}_1 +
  |c_2|^2 \rho_t^{(2)} \ \dot{\bf r}_2 ,
 \label{eq19}
\end{equation}
are properly defined in this limit; the former is a sum of partial
probability densities and the latter is the sum of the density
currents corresponding to each slit independently considered.
Moreover, in both cases these magnitudes are properly weighted with
the coefficients $|c_1|^2$ and $|c_2|^2$.
Thus, from a SQM point of view, an experiment with full decoherence
is equivalent to another one performed with each slit independently
open.
Something different happens, however, when this situation is studied
in terms of reduced quantum trajectories.
In this case, particles still move under the guidance of both partial
waves although interference has already disappeared.
This is because the damping describing interference suppression does
not account for the erasure of the information about the initial
presence of two slits.

It is important to note that within a purely BM approach the limit
discussed above has to be such that the corresponding Bohmian
trajectories will behave as unaware of the existence of a double--slit.
That is, particles started with initial conditions in one of the slits
will evolve with basically no information about the presence of the
other slit.
Since Bohmian trajectories cannot pass through the same point in
configuration space at the same time, such behavior can be explained
having in mind that the trajectories describing the
system--plus--environment interaction are embedded in a
3($N$+1)--dimensional configuration space.
In this sense, as the system--plus--environment interaction takes
place, the topology of these 3($N$+1)--dimensional trajectories will
be such that the projection of the system degrees of freedom onto the
system subspace will display crossings (at the same time).
Of course, this does not violate the single--valuedness condition of
BM, since crossings are only present in the projections.
Somehow this situation resembles what happens in classical mechanics,
where trajectories can pass through the same point of the configuration
space at the same time, but not in phase space.
Within our quantum--trajectory approach single--valuedness preserves;
once the dynamics of the environment degrees of freedom is not
considered explicitly, the single--valuedness condition directly
emerges in the system subspace.


\section{Decoherence in the double--slit experiment}
 \label{sec3}


\subsection{Model and simulation conditions}

The working model used here is based on the double--slit experiment
performed by Zeilinger {\it et al.} with cold neutrons \cite{Zeilinger1},
which has also been analyzed elsewhere by us from both an optical and a
SQM point of view \cite{Sanz3}.
Following the prescription given in Section~\ref{sec2}, our simulation
models the behavior of the neutron beam from the two slits to the
detector.
The double--slit arrangement has dimensions $a_1$--$d'$--$a_2$ =
21.9--104.1--22.5~$\mu$m (left slit/gap/right slit), and is at a
distance $L = 5$~m from the detector.
The wavelength of the incident neutron beam is $\lambda_{\rm dB} =
18.45$~\AA, corresponding to a subsonic velocity, $v = 214.4$~m/s.

Here interference is described by considering two Gaussian slits on
the $xy$--plane, with neutrons propagating along the $z$--direction.
As seen in~\cite{Sanz2,Sanz3}, Gaussian slits reproduce fairly well
the real experiment, avoiding at the same time single--slit diffraction
features \cite{Sanz1}.
To further simplify, we have assumed $\ell_y \gg \ell_x$, with $\ell_x$
and $\ell_y$ being the dimensions of the slits.
This allows us to neglect the motion along the $y$--direction due to
translational invariance (thus describing the particle motion only
along the $x$ and $z$ coordinates).
With this, $|\Psi^{(0)}\rangle$ in (\ref{eq4}) is given by a coherent
superposition of two Gaussian wave packets (here we consider that both
contribute equally, i.e., $c_1 = c_2 = 1/\sqrt{2}$), each one described
by
\begin{eqnarray}
 \psi_j (x,z) =
  \left( \frac{1}{2\pi\sigma_{x_j}\sigma_{z_j}} \right)^{1/2}
  {\rm e}^{-(x - x_j)^2/4 \sigma_{x_j}^2 + i p_{x_j} x/\hbar}
 \nonumber \\ \times \ \!
  {\rm e}^{-(z - z_j)^2/4 \sigma_{z_j}^2 + i p_{z_j} z/\hbar} ,
 \label{eq22}
\end{eqnarray}
with $j = 1, 2$.
Regarding the initial environment state, it is chosen as
$|E_0\rangle = | \ \! \mathbb{I} \ \! \rangle$, such that
$|\Psi\rangle=|\Psi^{(0)}\rangle \otimes | \ \! \mathbb{I} \ \!
\rangle$.

The time evolution of the (system) partial waves is carried out
numerically by Heller's method \cite{Sanz1}, which is exact in our
case because there is no external potential.
In the calculations, the Gaussian wave packets have been centered at
$x_{1,2} = (a_{1,2} \mp d)/2$ and $z_{1,2} = 0$, and incoherence has
been introduced by taking into account different propagation
velocities for each wave packet \cite{Sanz3}, given by
$p_{x_{1,2}}=\mp \hbar/a_{1,2}$ and $p_{z_{1,2}} = \sqrt{(2 \pi \hbar /
\lambda_{\rm dB})^2 - p_{x_{1,2}}^2}$ .
In order to minimize the spreading of the Gaussians along the
$z$--direction,
\begin{equation}
 \sigma_t^z = \sigma \,
  \sqrt{ 1 + \left( \frac{\hbar t}{m \sigma^2} \right)^2 } ,
 \label{eq23}
\end{equation}
during the time propagation, we have chosen $\sigma=2\bar{a}$ (with
$\bar{a}=(a_1+a_2)/2$) for both wave packets.
This ensures $\sigma_t^z \simeq \sigma$ for the whole propagation time.
As for the spreading along the $x$--direction, we have considered
$\sigma_{x_j} = a_j/4$, so that for $|x - x_j| = a_j/2$ the intensity
at the edge of slit $j$ amounts to
$|\psi_j (\pm a_j/2)|^2/|\psi_j (0)|^2 = e^{-2}$ (about 13.5\% of
the maximum value of the intensity, $|\psi_j (0)|^2$, reached when
$x = x_j$).
In this way, only a very small portion of the partial waves is out of
the boundaries defined by the edges of the slits.
This assumption is in good agreement with the error on the slit widths
experimentally reported in \cite{Zeilinger1}, according to which
neutrons penetrating through the boron wire (the physical gap between
the two slits) undergo a relatively strong attenuation.

According to Heller's propagation method, both partial waves are
evolved independently.
Then the parameter $\alpha_t$ is introduced whenever they are
superposed in order to obtain the intensity (\ref{eq9}).
The magnitude of $|\alpha|_t$ was empirically determined in \cite{Sanz3}
taking into account the coherence degree of the experimental results
($\Lambda_t = 0.632$) and the time--of--flight of neutrons ($t_f =
v/L = 2.33 \! \times \! 10^{-2}$~s), resulting a value of 0.36.
This value implies a coherence time $\tau_c = 2.26 \! \times \!
10^{-2}$~s, slightly smaller than the time--of--flight.
\begin{figure*}
 \sidecaption
 \epsfxsize=4.5in {\epsfbox{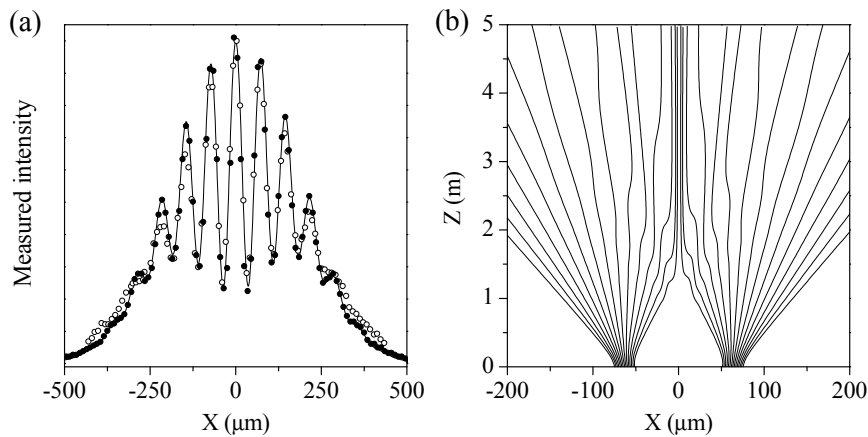}}
 \caption{\label{fig:1} (a) Comparison between experimental data
 ($\circ$) and the intensity obtained from quantum
 trajectory ($\bullet$) and SQM (full line) calculations for a
 double--slit experiment with cold neutrons \cite{Zeilinger1}.
 (b) Sample of trajectories illustrating the dynamics of the results
 shown in part (a).}
\end{figure*}

The reduced quantum trajectories were integrated according to equation
(\ref{eq17}) at the same time that the partial waves were propagated.
To obtain the statistical results, about 5,420 trajectories were used
in each calculation shown below, binning them in space intervals of
20~$\mu$m, which coincides with the experimental scanning slit width
\cite{Zeilinger1}.
These trajectories were initially distributed according to the
probability density $\rho^{(0)}$, thus ensuring the agreement with
SQM calculations through equation (\ref{eq13}).


\subsection{Numerical results}

In Figure \ref{fig:1}(a) the results obtained from the statistics of
trajectories ($\bullet$) are plotted together with the experimental
values ($\circ$).
Also to compare, we have included the results from SQM (solid line),
as given by equation (\ref{eq9}).
The excellent agreement between the experimental results and those
theoretically calculated by means of the reduced quantum trajectories
shows the suitability of the latter in describing decoherence in
interference phenomena.
The dynamical behavior of neutrons within this approach is illustrated
in Figure \ref{fig:1}(b), where a sample of trajectories associated to
the results in Figure \ref{fig:1}(a) is displayed.
From this plot it is apparent that, after the wave packets get close
enough (at a distance of 1 m from the two slits, approximately),
some trajectories (mainly those closer to the symmetry axis of the
experiment) begin to show the typical ``wiggling'' behavior
characterizing true Bohmian trajectories in interference processes
with no decoherence \cite{Sanz1}.
\begin{figure*}
 \sidecaption
 \epsfxsize=4.5in {\epsfbox{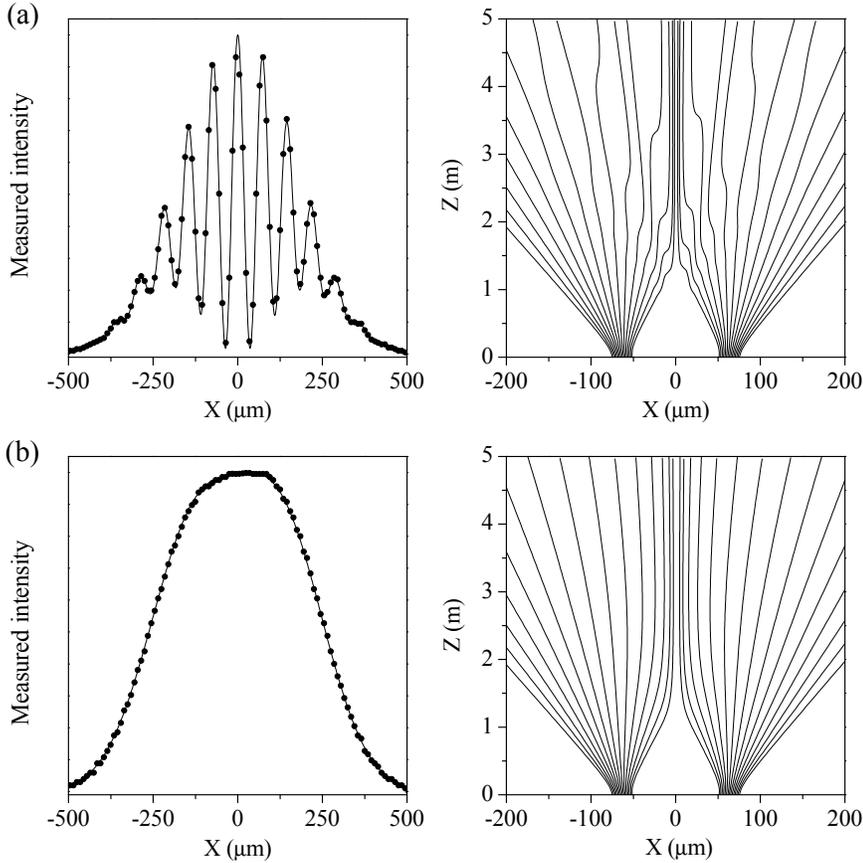}}
 \caption{\label{fig:2} Left: Intensity obtained from quantum
 trajectories ($\bullet$) and SQM (full line) for: (a) total coherence
 ($\tau_c = \infty$) and (b) null coherence ($\tau_c = 0$).
 Right: Samples of trajectories illustrating the dynamics of the
 results shown in the left part.}
\end{figure*}
Obviously, this behavior is more attenuated in both space and time
than in the case of true Bohmian trajectories (without decoherence)
because of the interference damping; in space because interference
effects are relevant only for the central channels, as can be seen in
Figure \ref{fig:1}(a), and in time because $t_f > \tau_c$.

In Figure \ref{fig:2} the two limit cases of coherence for this model
are illustrated: (a) total coherence ($\tau_c = \infty$) and (b) null
coherence ($\tau_c = 0$).
Similar to Figure \ref{fig:1}, the statistical results obtained by
means of reduced quantum trajectories and SQM (left) as well as a
sample of representative trajectories (right) are displayed.
Notice that, despite null coherence (see Figure \ref{fig:2}(b)), the
trajectories do not cross the symmetry axis that separates the regions
covered by each slit, as in the case of total coherence (see Figure
\ref{fig:2}(a)).
This is a manifestation of the contextual character of quantum
trajectories, which remains even under these conditions.
The absence of interference prevents the particles from undergoing the
typical ``wiggling'' motion that leads to the different diffraction
channels \cite{Sanz1}, but not from being non--locally correlated with
particles coming from the other slit.
Thus, within the approach proposed here we can see that decoherence
leads to a suppression of quantum interference, but not to loss of
memory on the initial context information (i.e., the existence of two
slits).
\begin{figure*}
 \sidecaption
 \epsfxsize=4.5in {\epsfbox{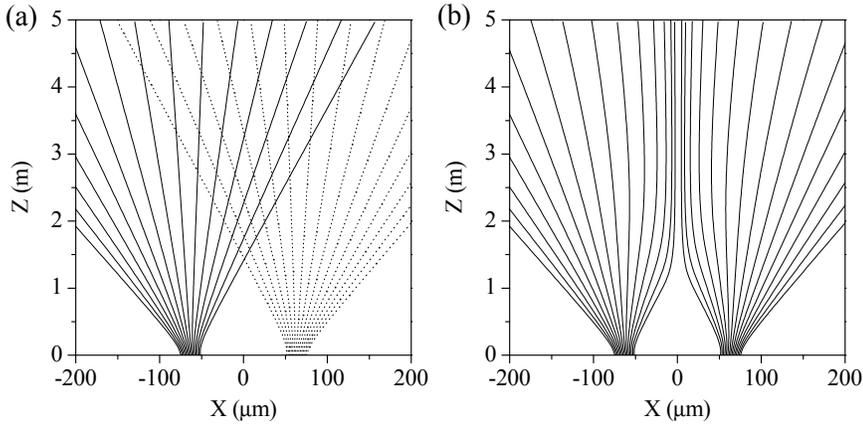}}
 \caption{\label{fig:3} Quantum trajectories for two different
 contexts: (a) slits independently open and (b) slits simultaneously
 open with $\tau_c = 0$.}
\end{figure*}
This is somehow similar to what happens in BM when trying to reach
the classical limit without appealing to any decoherence mechanism
\cite{Sanz4}; classical--like statistical patterns emerge, but
contextuality does not disappear.

According to the preceding statement, the structure of the reduced
quantum trajectories allows to characterize different situations by
their contextuality.
That is, a situation where two slits are independently open can be
easily distinguished from another where both are simultaneously open
but there is total decoherence.
These cases are illustrated in Figure \ref{fig:3}.
Although the trajectories started close to the outermost edges of each
slit are identical in both cases, as the initial positions approach the
innermost edges the behavior of the trajectories gets different.
When the slits are independently open (see Figure \ref{fig:3}(a)), each
set of trajectories is associated to an independent wave, and the
crossing between trajectories coming from different slits is allowed
because their dynamics are totally uncoupled.
On the contrary, when the two slits are simultaneously open (see Figure
\ref{fig:3}(b)), the dynamics are still strongly coupled, leading to an
apparent ``repulsion'' between both sets of trajectories as they meet
at about $z \approx 1$~m.
Note that this effect can only be detected by means of the
trajectories, since the measured intensity does not reveal any clue
about it; in both cases it is a sum of the probabilities associated
to each slit, as given by equation (\ref{rhodec}).
Of course, true Bohmian trajectories would show that this non--crossing
takes place in the 3($N$+1)--dimensional configuration space where the
system--plus--environment is described.


\section{Conclusions}
 \label{sec4}

Realistic simulation of system--plus--environment interactions from
purely quantum--mechanical approaches constitutes a hard computational
task due to the many degrees of freedom involved.
Nonetheless, it has become a topic of mayor interest in recent years
\cite{chapter} because of its important in different fields in physics,
chemistry, and biology.
Here we have proposed a quantum trajectory description, based on BM,
that allows to study such problems without taking into account
explicitly the evolution of the environment degrees of freedom.
For that, the trajectories are directly obtained from the system
reduced density matrix, with the action of the environment arising
from a damping term that appears in such reduced density matrix.
In this way, these trajectories reduce computational efforts as well
as provide a physical inside on the physics taking place in phenomena
where the system coherence is lost because of environment effects.

To illustrate the feasibility of our trajectory approach, we have
applied it to the problem of decoherence in interference phenomena, in
particular, to the disappearance of interferences in the double--slit
experiment.
With this model it is shown that, effectively, interference fringes
disappear, although the trajectory dynamics is still influenced by both
slits, despite what one would expect.
This is because the trajectory dynamics is constrained to the system
reduced subspace instead of the full system--plus--environment
configuration space, from which one can see that the projection of
the system degrees of freedom onto the system subspace violates the
non--crossing property of BM.
Furthermore, based on this fact, we have also shown that the reduced
trajectories can be used to distinguish between experiments that give
identical SQM results.
This is the case, for example, of an experiment performed with each
slit independently open and another with both slits simultaneously
open but total decoherence.


\section*{Acknowledgements}

This work has been supported by Ministerio de Educaci\'on y Ciencia
(Spain) under projects MTM2006--15533, CONSOLIDER 2006--32, and
FIS2004--02461; Comunidad de Madrid under the project S--0505/ESP--0158;
and Agencia Espa\~nola de Cooperaci\'on Internacional under the project
A/6072/06.
A.S. Sanz would also like to thank the Ministerio de Educaci\'on y
Ciencia for a ``Juan de la Cierva'' Contract.


\end{document}